\documentclass[12pt,preprint]{aastex}
\usepackage{epsf}
\usepackage{graphicx}
\usepackage{lscape}
\usepackage{natbib}

\tabletypesize{\scriptsize}

\newcommand{\etal}{{et al.}\ }

\def\gappeq{\mathrel{ \rlap{\raise.5ex\hbox{$>$}}
                      {\lower.5ex\hbox{$\sim$}}  } }

\begin{document}
\shorttitle{51 Oph Observations with the Keck Nuller}
\shortauthors{Stark \etal}


\title{51 Ophiuchus: A Possible Beta Pictoris Analog Measured with the Keck Interferometer Nuller}

\author{Christopher C. Stark\altaffilmark{1}, Marc J. Kuchner\altaffilmark{2}, Wesley A. Traub\altaffilmark{3}, John D. Monnier\altaffilmark{4}, Eugene Serabyn\altaffilmark{3}, Mark Colavita\altaffilmark{3}, Chris Koresko\altaffilmark{5}, Bertrand Mennesson\altaffilmark{3}, Luke D. Keller\altaffilmark{6}}

\altaffiltext{1}{Department of Physics, University of Maryland, Box 197, 082 Regents Drive,
College Park, MD 20742-4111, USA;
starkc@umd.edu}
\altaffiltext{2}{NASA Goddard Space Flight Center, Exoplanets and Stellar
Astrophysics Laboratory, Code 667, Greenbelt, MD 20771}
\altaffiltext{3}{Jet Propulsion Laboratory, California Institute of Technology, Pasadena, CA 91109}
\altaffiltext{4}{Department of Astronomy, University of Michigan, 500 Church Street, Ann Arbor, MI 48109}
\altaffiltext{5}{Argon ST, Fairfax, VA 22033}
\altaffiltext{6}{Department of Physics, Ithaca College, Ithaca, NY 14850}

\begin{abstract}

We present observations of the 51 Ophiuchi circumstellar disk made with the Keck interferometer operating in nulling mode at N-band.  We model these data simultaneously with VLTI-MIDI visibility data and a \emph{Spitzer} IRS spectrum using a variety of optically-thin dust cloud models and an edge-on optically-thick disk model.  We find that single-component optically-thin disk models and optically-thick disk models are inadequate to reproduce the observations, but an optically-thin two-component disk model can reproduce all of the major spectral and interferometric features.  Our preferred disk model consists of an inner disk of blackbody grains extending to $\sim 4$ AU and an outer disk of small silicate grains extending out to $\sim 1200$ AU.  Our model is consistent with an inner ``birth" disk of continually colliding parent bodies producing an extended envelope of ejected small grains.  This picture resembles the disks around Vega, AU Microscopii, and $\beta$ Pictoris, supporting the idea that 51 Ophiuchius may be a $\beta$ Pictoris analog.


\end{abstract}


\section{Introduction}

51 Ophiuchi (51 Oph), a rapidly rotating B9.5Ve star located at $131^{+17}_{-13}$ pc \citep{p97}, shows an infrared (IR) excess ($L_{\rm IR} / L_{\star} \approx 2$\%) in its spectral energy distribution (SED) due to the presence of silicate grains \citep{fatk93, m01, k08}.  This system probably represents a rare nearby example of a young debris disk with gas, a planetary system just entering the late stages of formation, after the primordial gas has substantially dissipated, but before terrestrial planets have finished accreting.  Its nearly edge-on disk of gas and dust and its variable absorption features \citep{gs93, r02} suggests that 51 Oph may be an analog of $\beta$ Pictoris ($\beta$ Pic).

Several spectroscopic observations support the presence of an edge-on gaseous disk around 51 Oph.  Double-peaked H$\alpha$ emission marks the presence of ionized hydrogen gas in Keplerian orbit \citep{l04}.  Sharp CO bandhead emission features rule out a spherically symmetric wind and reveal a gaseous disk with temperature ranging from 2000 -- 4000 K interior to the dust sublimation radius \citep{t05}.  The CO bandhead observations also point to a disk with inclination $i = 88^{+2}_{-35} \!^\circ$ \citep{t05} or $i = 83^{+7}_{-0}\,^\circ$ \citep{b07}.  A spectral line analysis performed by \citet{dbr97} revealed a large projected rotational velocity for the star of $v \sin{i} = 267 \pm 5$ km s$^{-1}$.  Gas absorption features observed by \citet{gs93} and \citet{r02} are also consistent with an edge-on disk.  


The spatial structure of the 51 Oph dust disk remains puzzling.  An HST ACS V-band non-detection \citep{d07} and a Keck 18 $\mu\rm{m}$ non-detection \citep{j01} suggest that the disk is relatively compact.  However, far-IR photometry reveals cold dust grains extending out to thousands of AU \citep{wcg88, lde97a}.

\citet{l04} obtained the first spatially-resolved measurements of the 51 Oph disk with the Mid-IR Interferometric (MIDI) instrument on the Very Large Telescope Interferometer (VLTI).  The large visibility values they measured ($\sim 0.65$) imply that the 51 Oph disk is relatively compact along the VLTI-MIDI projected baseline (101.2 m, 23$^{\circ}$ E of N).  \citet{l04} fit the VLTI-MIDI visibility with a Gaussian disk and found the FWHM to be 7 mas, or 0.9 AU, at 12.5 $\mu\rm{m}$.

Here we present new spatially-resolved observations of 51 Oph using the Keck Interferometer Nuller that help to constrain the geometry of the dust disk.  We compare these to the VLTI-MIDI observations \citep{l04} and \emph{Spitzer} IRS observations \citep{k08}.  We simultaneously model all three data sets using two simple, edge-on disk models: an optically-thin model based on our zodiacal cloud and a two-layer model based on the \citet{cg97} disk model.

\section{Observations}


Observations of 51 Oph were conducted using the twin Keck telescopes atop Mauna Kea, Hawaii, operated in nulling interferometric mode on 2 Apr 2007.  51 Oph was observed twice in the N-band (8 - 13 $\mu\rm{m}$) at an hour angle of $\approx 0.5$, with a projected baseline of 66.2 m at a position angle of $47^{\circ}$.  A calibrator star, Epsilon Ophiuchi (HD146791), was observed once following the target observations.  Table \ref{observationstable} lists the details of our observations.

The Keck Nuller operates with each Keck telescope aperture divided into two sub-apertures for a total of four apertures \citep[see][for details]{s05, c08}.  Long-baseline fringes formed by combining the light from opposite telescopes are used to null the light from the central star and observe any spatially resolved structures.  Short-baseline fringes formed by combining the light from two neighboring sub-apertures are used to remove the thermal background.

The observable quantity is the ``null leakage," the ratio of the amplitude of the short baseline fringe with the long baseline null fringe on target to the amplitude of the short baseline fringe with the long baseline constructive fringe on target \citep[see][for details]{s05, b08}.  We estimated and subtracted the systematic null leakage by comparing the measured null leakage of the calibration star, $\epsilon$ Oph, with the expected null leakage for a limb-darkened star with the same diameter.  We estimated the diameter of $\epsilon$ Oph as $2.94$ mas and adopted 1.5 mas error bars---much larger than the true size error bars---as a simple means of enforcing a systematic noise floor based on our estimate of the instrument performance.


\section{Data \& Analysis}

\subsection{Keck Nuller null leakage}

\begin{figure}
\begin{center}
\includegraphics[width=6.5in]{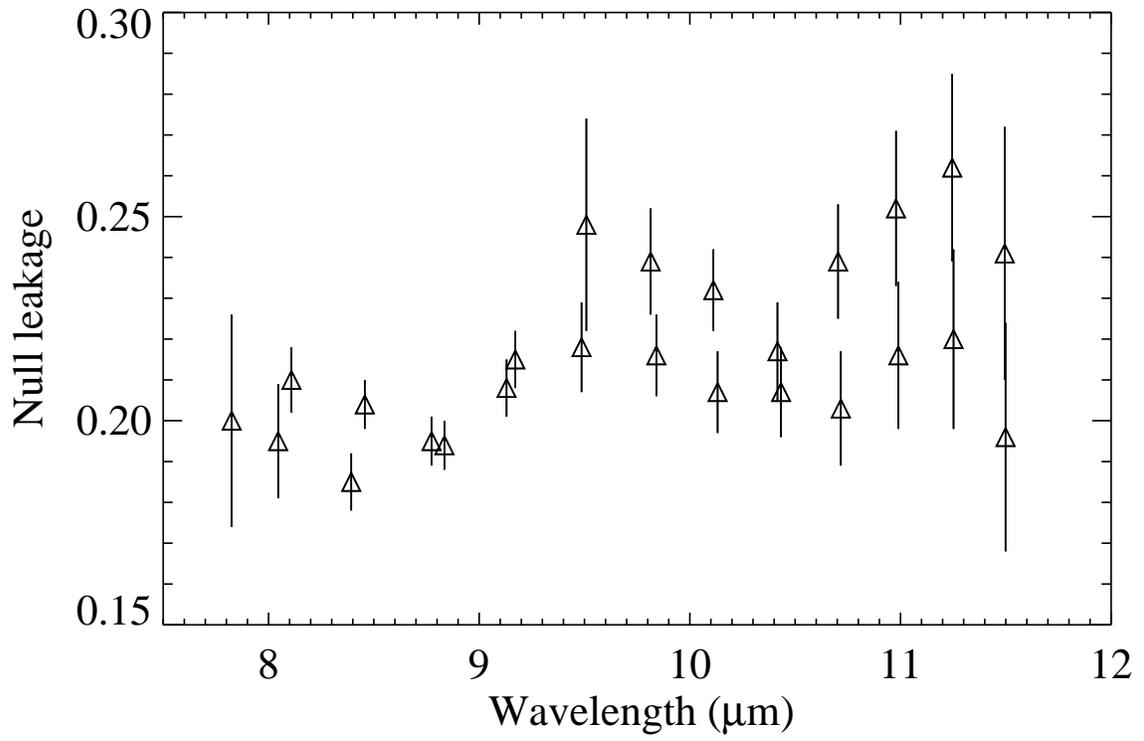}
\caption{Keck Nuller null leakage measurements for 51 Oph. \label{kecknulldatafig}}
\end{center}
\end{figure}

Figure \ref{kecknulldatafig} presents the calibrated null leakage for 51 Oph.  We combined the data from both observations, which had nearly identical projected baselines.  We limited our analyses to the 8 -- 11.6 $\mu\rm{m}$ range since noise from decreased instrument throughput rendered data beyond 11.6 $\mu\rm{m}$ unusable.  For wavelengths less than 11.6 $\mu\rm{m}$, the null leakage remains relatively flat with a $\sim$1--$\sigma$ rise near 9.8 $\mu\rm{m}$.

We first modeled the null leakage at two different wavelengths with uniform disk and Gaussian surface brightness profiles.  We found angular diameters of $13.5 \pm 0.5$ mas and $18.5 \pm 0.4$ mas at 8 and 10 $\mu\rm{m}$, respectively, for the uniform disk profile.  For a Gaussian surface brightness profile, we found FWHM of $8.1 \pm 0.3$ mas and $11.3 \pm 0.2$ mas at 8 and 10 $\mu\rm{m}$, respectively.  At a distance of 131 pc, 10 mas corresponds to a transverse distance of 1.3 AU, suggesting that the disk is either truncated at a close circumstellar distance, or the axis of rotation of the near edge-on disk is oriented within a few tens of degrees of the projected Keck baseline. 

To better understand the geometry of the 51 Oph system, we examined our Keck Nuller observations together with the observations of 51 Oph made with VLTI-MIDI and \emph{Spitzer}.  Figure \ref{bestfit2comp_noshortlambda_figure} shows a collection of three independent data sets from observations of 51 Oph: the \emph{Spitzer} IRS spectrum \citep{k08} in the top panel, our N-band Keck Nuller null leakage in the middle panel, and the N-band VLTI-MIDI visibility data \citep{l04} in the bottom panel.  

\subsection{MIDI visibility}

\begin{figure}
\begin{center}
\includegraphics[width=6.in]{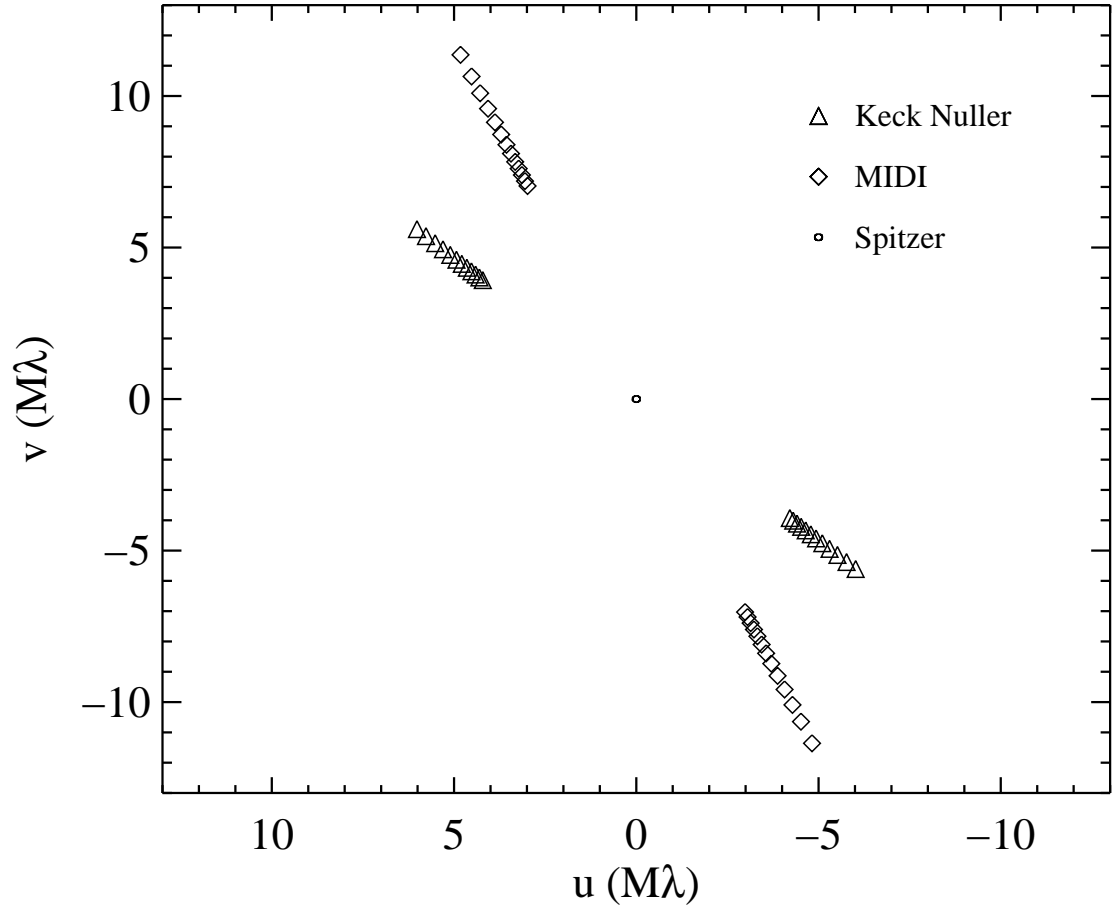}
\caption{$(u,v)$ coverage of the observations discussed in this paper.  The outer boundary of the \emph{Spitzer} $(u,v)$ coverage is shown as a circle, corresponding to the \emph{Spitzer} mirror diameter at 8 microns.\label{uvcoverage_figure}}
\end{center}
\end{figure}

The VLTI-MIDI visibility was obtained on 15 June 2003 with a projected baseline of 101.2 m at a position angle of $23^{\circ}$ \citep{l04}.  Figure \ref{uvcoverage_figure} shows how incorporating this data set improves the $(u,v)$ coverage of our analysis.  Although the VLTI-MIDI baseline was oriented within $25^{\circ}$ of the Keck baseline, modeled uniform disk and Gaussian surface-brightness profile sizes are approximately 35\% smaller for VLTI-MIDI measurements than for Keck Nuller measurements.  When we modeled the VLTI-MIDI measurements using a uniform surface-brightness disk, we found best-fit angular diameters of $8.5 \pm 1.4$ mas at 8 $\mu\rm{m}$ and $12.4 \pm 1.1$ mas at 10 $\mu\rm{m}$, consistent with \citet{l04}.  For a Gaussian model, we found a FWHM of $5.0 \pm 0.9$ mas at 8 $\mu\rm{m}$ and a FWHM of $7.7 \pm 0.6$ mas at 10 $\mu\rm{m}$.  

The middle panel in Figure \ref{bestfit2comp_noshortlambda_figure} shows a 1--$\sigma$ rise at 9.8 $\mu\rm{m}$ in the Keck null leakage.  The VLTI-MIDI visibility contains a 1--$\sigma$ dip at 9.7 $\mu\rm{m}$.  These features mirror the 10 $\mu\rm{m}$ silicate emission feature shown in the \emph{Spitzer} IRS spectrum (Section \ref{spitzer_data}) and suggest that 51 Oph is more extended near 10 $\mu\rm{m}$.

\subsection{Spitzer IRS spectrum\label{spitzer_data}}

The mid-infrared spectrum, shown in the top panel of Figure \ref{bestfit2comp_noshortlambda_figure}, was obtained on 22 March 2004 using the Infrared Spectrograph (IRS) on the \emph{Spitzer Space Telescope} \citep{k08}.  \emph{Spitzer} observed 51 Oph in staring mode from 5 to 36 $\mu\rm{m}$ using the Short-Low (SL) module from 5 to 14 $\mu\rm{m}$, the Short-High (SH) module from 10 to 19 $\mu\rm{m}$, and the Long-High (LH) module from 19 to 36 $\mu\rm{m}$.  SL has a resolving power of R$=60$--128, while SH and LH have R$\sim$600.  51 Oph was observed using three slit positions stepped across the nominal position of the source.  The spectral extraction and calibration
methods are described in \citet{k08}.  Figure \ref{bestfit2comp_noshortlambda_figure} shows that the spectrum exhibits a pronounced 10 $\mu\rm{m}$ silicate feature and a small 18 $\mu\rm{m}$ silicate feature.  

Emission from polycyclic aromatic hydrocarbon (PAH) molecules contributes to the mid-IR spectra of many Herbig Ae/Be stars.  We considered the possibility that this emission could contribute to our data on 51 Oph.  \citet{k08} included 51 Oph in a study that looked for correlations between disk structure and mid-IR PAH emission from intermediate-mass young stars.  They found no measurable emission from PAH features in the 6--13 $\mu$m range in the mid-IR spectrum of 51 Oph.

\section{Modeling the 51 Oph disk}

There have been several previous attempts to model the 51 Oph IR excess.  \citet{wcg88} fit IRAS photometric data with an optically thin, spherically symmetric dust shell model and found a best fit model with a dust density proportional to $r^{-1.3}$ and dust temperatures ranging from $\sim 100$ K to $\sim 1000$ K.  \citet{fatk93} modeled photometric data from the Infrared Telescope Facility (IRTF) and estimated that the IR excess could be attributed to astronomical silicates smaller than 5 $\mu\rm{m}$, ranging in temperature from 400 K to 1000 K.  \citet{l04} compared an optically-thick Herbig Ae disk model with a puffed inner rim \citep{d01} to the 51 Oph MIDI visibility in the 8 -- 14 $\mu\rm{m}$ range and IR ISO spectra and found that such a model fit poorly.

We developed new models for 51 Oph to incorporate the new data from \emph{Spitzer} and the Keck nuller.  We do not model the detailed mineralogy of the 51 Oph disk and do not intend for our models to explain all of the observed spectral features to high numerical accuracy.  We focus on the 3-D dust density distribution and disk models that qualitatively describe all three data sets.

We adopted the following fitting procedure to simultaneously fit all three data sets:
\begin{enumerate}
	\item Model the disk contribution to the \emph{Spitzer} IRS spectrum to obtain the radial distribution of grains, the grain size, and the surface density for each model component.
	\item Using the parameters determined by the best-fit \emph{Spitzer} IRS spectrum, fit the interferometric data using 3-D models of the dust distribution  to obtain the scale height of each component and the disk position angle.
\end{enumerate}

To model the disk contribution to the \emph{Spitzer} IRS spectrum, we first calculated the IR excess.  We modeled the stellar contribution to the \emph{Spitzer} IRS spectrum as an ideal blackbody with an effective temperature of 10,000 K and a luminosity of 260 $\rm{L_\sun}$ at 131 pc \citep{vdA01}.  The stellar continuum contributes on the order of a few percent in the N-band, so any uncertainties in the stellar luminosity or temperature are too small to significantly impact the interferometric or spectral responses of our models.  For example, a $10\%$ increase in the assumed luminosity would only raise the stellar fractional flux contribution from $\sim 5\%$ to $\sim 7\%$ at 10 $\mu\rm{m}$.

The 51 Oph IR excess exhibits a sharp increase in flux at wavelengths shorter than 7.5 $\mu\rm{m}$ that resembles a continuum source.  However, this feature may not be continuum.  Using ISO-SWS spectra with a spectral resolution of $\sim .02\%$, \citet{vdA01} showed this region exhibits emission features from hot circumstellar gas species, including $\rm{H_2O}$, NO, CO, and $\rm{CO_2}$.  In light of this contamination, we chose to ignore the spectrum at wavelengths shorter than 7.5 $\mu\rm{m}$ while we were fitting our dust cloud models.  We also resampled the \emph{Spitzer} IRS data to a resolution of $\Delta \lambda / \lambda = 0.0185$ to conserve computer time.

To create 3-D optically-thin disk models, we used the ZODIPIC software package \citep{mkh04}.  This software (available online at http://asd.gsfc.nasa.gov/Marc.Kuchner/home.html) synthesizes images of optically thin dust disks based on the \citet{k98} empirical model of the solar zodiacal cloud as observed by COBE DIRBE.  The grain size and inner and outer radius of each modeled disk component were determined by the best fit to the \emph{Spitzer} IRS spectrum.

\subsection{Single-component optically-thin models}

For our first models we used a single component of silicate dust grains of size $s$ distributed from an inner radius, $r_{\rm inner}$, to an outer radius, $r_{\rm outer}$.  We assumed a density distribution similar to that of the zodiacal cloud \citep{k98}; the surface density $\Sigma(r) \propto r^{-0.34}$.  We numerically calculated the temperature as a function of circumstellar distance based on the stellar spectrum and dust optical constants, accounting for the heating of small grains above the local blackbody temperature.  We examined 100 grain sizes, $s$, ranging from 0.05 -- 2.5 $\mu\rm{m}$ and used the astronomical silicate emissivities from \citet{dl84}.  We used a non-linear least squares fitting routine to determine the best fit inner and outer disk radii for each grain size.

\begin{figure}
\begin{center}
\includegraphics[width=4.in]{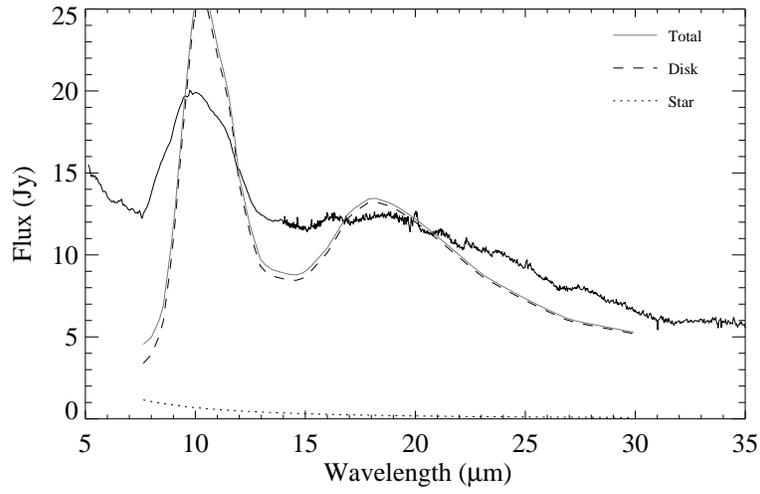}
\caption{Best fit single-component disk model to the \emph{Spitzer} spectrum (solid black line).  The best fit features 1 $\mu\rm{m}$ grains, distributed from 0.65 AU to 189 AU.\label{bestfit_singlecomponent_figure}}
\end{center}
\end{figure}

The best-fit single-component disk model, shown in Figure \ref{bestfit_singlecomponent_figure}, consists of 1.0 $\mu\rm{m}$ grains distributed from the grain sublimation radius ($\approx 0.65$ AU) to 189 AU.  As you might imagine, this single component does a poor job of fitting the complexities of the \emph{Spitzer} IRS spectrum.  We found that a single-component model can not adequately reproduce the relative 10 $\mu$m to 18 $\mu\rm{m}$ flux ratio, the width of the 10 $\mu\rm{m}$ feature, and the flux in the 13 -- 15 $\mu\rm{m}$ range.

\subsection{Two-component optically-thin models}

An additional disk component, either of different composition, grain size, or location, appears necessary to fit the \emph{Spitzer} IRS spectrum.  Therefore, we examined several models consisting of two optically-thin components: an outer disk of small dust grains ($\lesssim 5\; \mu\rm{m}$), which contribute the 10 $\mu\rm{m}$ silicate emission feature, and an inner disk of large blackbody grains.  Our optically-thin models are defined by 10 adjustable parameters: inner and outer radii of the outer disk ($r_{\rm in,1}$, $r_{\rm out,1}$), inner and outer radii of the inner disk ($r_{\rm in,2}$, $r_{\rm out,2}$), grain size in the outer disk ($s_{\rm 1}$), scale height of the outer disk ($h_{\rm 1}/r$), scale height of the inner disk ($h_{\rm 2}/r$), surface density scaling factor of the outer disk ($\Sigma_{\rm 1}$), surface density scaling factor of the inner disk ($\Sigma_{\rm 2}$), and a common position angle for both disk components (PA).  We assumed a fixed inner disk grain size, $s_{\rm 2} = 100\; \mu\rm{m}$, a fixed dust density radial distribution, $n(r) \propto r^{-1.34}$ (see below), and a fixed disk inclination, $i = 90^{\circ}$.

First, we fit our two-component model to our resampled \emph{Spitzer} IRS spectrum.  We examined 20 values for $s_1$ ranging from $0.1$ -- $3\; \mu\rm{m}$ and explored a wide range of  inner and outer disk radii ranging from the dust sublimation radius for silicate grains (0.65 AU for 1 $\mu\rm{m}$ grains) to thousands of AU.  Generally, for the models that best fit the spectrum, the large grain component stretched from $\sim 0.6$ AU out to $\sim 5$ AU and the small grain component stretched from a few AU out to $\sim 1000$ AU.

\begin{figure}
\begin{center}
\includegraphics[width=4.in]{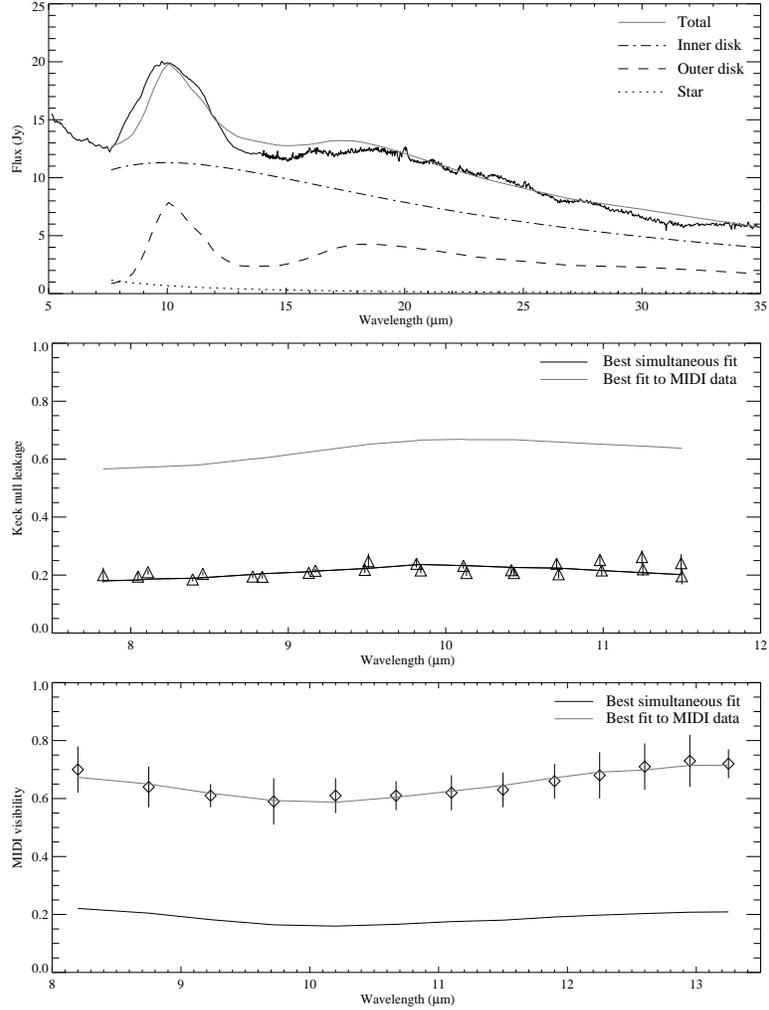}
\caption{Model A, the best fit to the three data sets using a 2-component optically thin disk model.  {\it Top panel:} the observed \emph{Spitzer} IRS spectrum (solid black line), and best-fit modeled spectrum.  {\it Middle panel:} the observed Keck Nuller null leakage (triangles) shown with error bars, the best-fit modeled null leakage when simultaneously fitting both Keck and VLTI-MIDI data sets (black line), and the best-fit null leakage when fitting the VLTI-MIDI data set alone (gray line). {\it Bottom panel:} the observed VLTI-MIDI visibility (diamonds) shown with error bars and modeled VLTI-MIDI visibilities.  The parameters of Model A which best fit the two interferometric data sets simultaneously are listed in Table \ref{bestfitstable}.  The best fit to the VLTI-MIDI data alone is shown in gray and corresponds to a disk with a position angle of $117^{\circ}$, an outer disk scale height of $h_1/r \approx 0.038$, and an inner disk scale height of $h_2/r \approx 0.024$.\label{bestfit2comp_noshortlambda_figure}}
\end{center}
\end{figure}

The top panel in Figure \ref{bestfit2comp_noshortlambda_figure} shows the total flux of the best fit model, which we refer to as ``Model A."  It also shows the contribution of each component of this model to the total flux.  As shown in Figure \ref{bestfit2comp_noshortlambda_figure}, Model A qualitatively reproduces all of the major features in the \emph{Spitzer} IRS spectrum.

We used the parameters that best fit the \emph{Spitzer} IRS spectrum to create 3-D optically-thin disk models using ZODIPIC.  We assumed a fixed disk inclination of $90^\circ$.  We examined 237 values of disk scale heights ranging from $h/r = 0.007$ to $h/r = 0.2$ for both the inner and outer disks.  We examined position angles from 0 to 180$^{\circ}$ in 1$^{\circ}$ increments.  We calculated the Keck Nuller null leakage and VLTI-MIDI visibility for all combinations of disk scale heights and position angle to find the best-fit Model A, the parameters for which are listed in Table \ref{bestfitstable}.  To calculate the Keck Nuller null leakage and VLTI-MIDI visibility, we used a software suite that we designed to model these instruments \citep[see][]{b08}.  We confirmed that the results from our software suite agreed with the Visibility Modeling Tool, a tool developed by the NExSci for simulating KIN data (http://nexsciweb.ipac.caltech.edu/vmt/vmtWeb/).

The interferometric responses of Model A are shown in the middle and bottom panels of Figure \ref{bestfit2comp_noshortlambda_figure}.  The best simultaneous fit to the Keck Nuller null leakage and VLTI-MIDI visibility, shown in black, illustrates that Model A does not satisfactorily reproduce both interferometric responses simultaneously.  The Keck null leakage is well-fit, but the VLTI-MIDI visibility is underestimated by a factor of $\sim 3$, indicating that this dust disk model is too extended along the VLTI-MIDI baseline.  Figure \ref{bestfit2comp_noshortlambda_figure} also shows in gray the response corresponding to the best fit to the MIDI data alone, which does not produce a satisfactory Keck null leakage.

\begin{figure}
\begin{center}
\includegraphics[width=4.in]{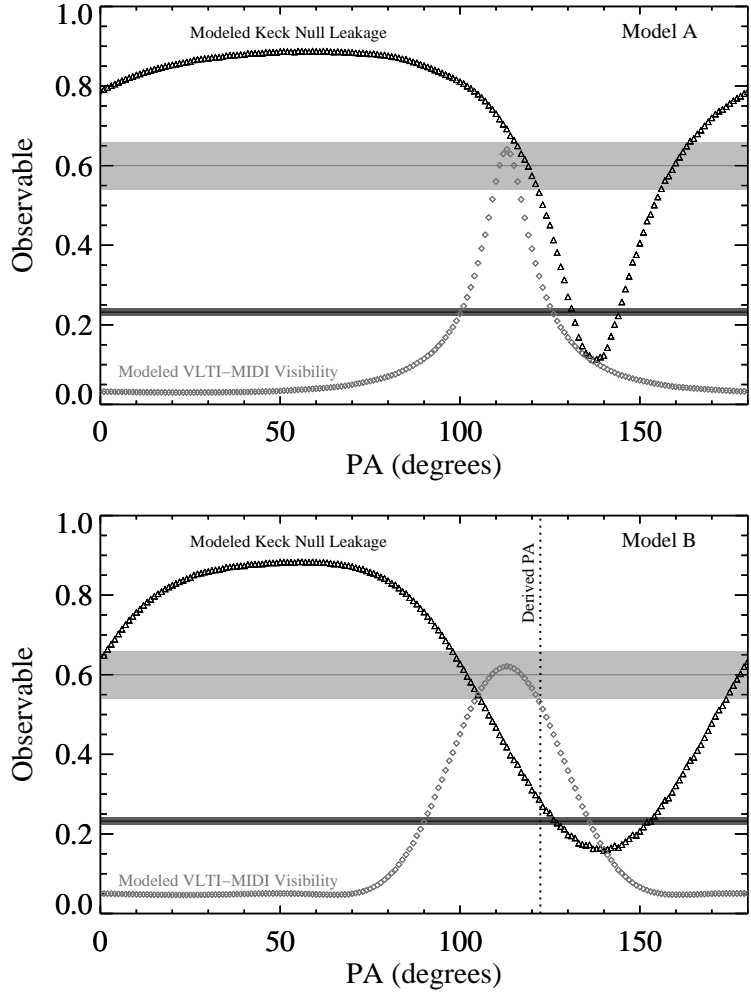}
\caption{\emph{Top panel:} Modeled interferometric response at 10 $\mu\rm{m}$ as a function of position angle for the Keck Nuller (black triangles) and VLTI-MIDI (gray diamonds) for Model A.  The solid black and gray lines show the measured 10 $\mu\rm{m}$ responses for Keck and VLTI-MIDI, respectively, along with shaded regions representing their respective measurement uncertainties.  The modeled Keck and VLTI-MIDI responses do not cross their measured values together at any one position angle.  \emph{Bottom panel:} Modeled interferometric response at 10 $\mu\rm{m}$ for Model B.   The modeled responses cross the corresponding measured values at approximately 122$^{\circ}$.\label{response_vs_pa_figure}}
\end{center}
\end{figure}

Figure \ref{response_vs_pa_figure} explains the source of the problem.  The top panel in Figure \ref{response_vs_pa_figure} shows the Keck and VLTI-MIDI 10 $\mu\rm{m}$ responses for Model A as a function of disk position angle.  The VLTI-MIDI response of Model A (gray diamonds) crosses the measured VLTI-MIDI value (gray line) at a position angle of $\sim 115^{\circ}$.  The Keck Nuller response of Model A (black triangles) crosses the measured Keck null leakage (black line) at a position angle of $\sim 131^{\circ}$.  To fit both data sets, the 10 $\mu\rm{m}$ responses for the model would need to intersect their respective measured values at a single position angle.  The top panel in this figure clearly shows that this does not happen; there is no single position angle for the model that works for both the Keck and VLTI-MIDI data.

Figure \ref{response_vs_pa_figure} also reveals how a new model can remedy the problem.  The position angles of the maximum in the VLTI-MIDI response and the minimum in the Keck response are fixed and correspond to the alignment of the instrument's projected baseline with the disk axis of rotation.  So to improve our fit to the interferometric data, we must change the model so that we broaden the widths of the maximum in the VLTI-MIDI response and minimum in the Keck response; we must make the model more azimuthally symmetric.

We attempted to accomplish this improvement manually by increasing the scale height of the disk models.  However, making this change alone reduces the maximum VLTI-MIDI visibility of the model until it no longer reaches the measured value.  So to compensate, we also reduced the outer radius of the inner disk; we forced the inner and outer radii of the inner disk to 0.54 AU and 4 AU, respectively, and re-ran the fitting procedure.  We call the resulting best-fit ``Model B."  The bottom panel in Figure \ref{response_vs_pa_figure} shows that the 10 $\mu\rm{m}$ interferometric responses  of Model B cross the measured values approximately simultaneously at a single position angle of 122$^{\circ}$, indicated with a dotted vertical line in the figure.

Although Model B does not fit the \emph{Spitzer} IRS spectrum as well as Model A, it still qualitatively reproduces the spectrum's major features, as shown in the top panel in Figure \ref{bestfit2comp_noshortlambda_figure_7}.  The middle and bottom panels of Figure \ref{bestfit2comp_noshortlambda_figure_7} show that these changes significantly improve the fit to the interferometric data; Model B is consistent with both the Keck null leakage and VLTI-MIDI visibility.

\begin{figure}
\begin{center}
\includegraphics[width=6.in]{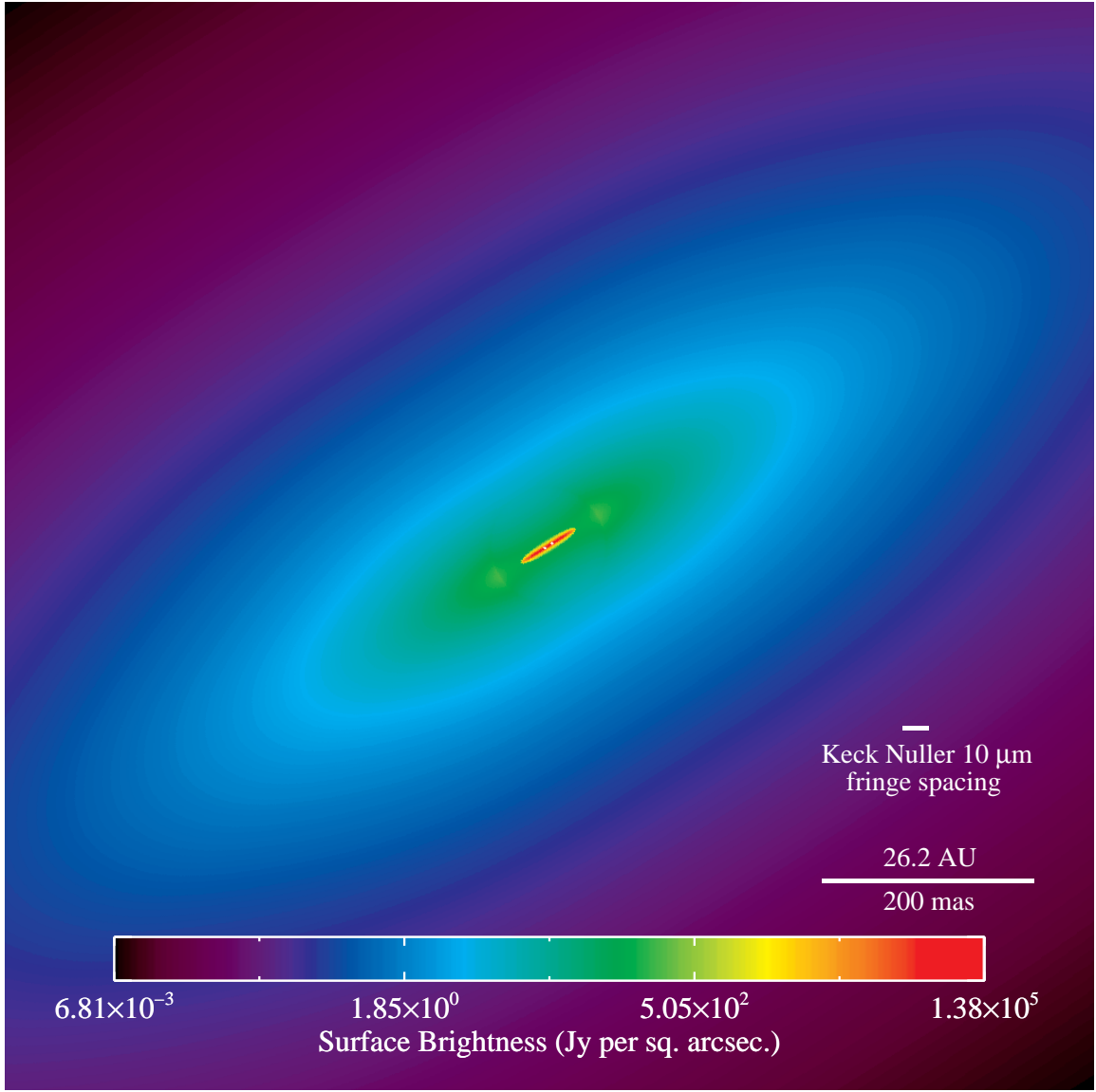}
\caption{Simulated 10 $\mu\rm{m}$ image of Model B in units of flux per pixel, with a pixel size of 1 mas.  The disk midplane is oriented at 122$^{\circ}$ East of North.  The parameters of Model B are listed in Table \ref{bestfitstable}.  \label{image_figure}}
\end{center}
\end{figure}

Figure \ref{image_figure} shows a simulated image of our Model B circumstellar disk at 10 $\mu\rm{m}$ with 1 mas pixel resolution.  The inner disk of large dust grains, truncated at 30.5 mas (4 AU), is the brightest feature of our model.  The outer disk extends from 53 mas to $\sim$9200 mas, beyond the range of the figure.

Our models imply that 51 Oph hosts a cloud of small grains located at hundreds to thousands of AU from the star, supporting the models of \citet{wcg88} and \citet{lde97a}.  Our models are consistent with limits placed on the disk flux by previous non-detections.  An HST ACS non-detection at V-band using the $1.8''$ occulting spot limits the disk surface brightness to $3 \times 10^{-3}$ Jy arcsec$^{-2}$ between $2''$ and $4''$ \citep{d07} and a Keck 18 $\mu\rm{m}$ nondetection places an upper limit on the surface brightness at $1''$ of 2\% of the peak flux \citep{j01}.  Our Model A, which extends to $\sim 1200$ AU, has a mean V-band surface brightness of $8 \times 10^{-5}$ Jy arcsec$^{-2}$ between $2''$ and $4"$ and a mean 18 $\mu\rm{m}$ surface brightness at $1''$ of 0.2\% of the peak flux, well within the ACS and Keck non-detection limits.  Our Model B has a mean V-band surface brightness of $1.4 \times 10^{-4}$ Jy arcsec$^{-2}$ between $2''$ and $4''$ and a mean 18 $\mu\rm{m}$ surface brightness at $1''$ of 0.4\% of the peak flux, also well within the non-detection limits.  We also compared a 10.6 $\mu\rm{m}$ model image against recent diffraction-limited Keck imaging using segment tilting interferometry on a single telescope \citep{m09}.  Our model appears consistent with these data, which indicate the observed characteristic emission scale to be $<$30 mas, with $>$95\% of emission arising within an aperture of 1.5".

\citet{t08} observed the inner portions of the 51 Oph disk at K-band with VLTI-AMBER.  \citet{t08} found the best fit to the 2.2 micron continuum visibility measurements using a narrow ring 0.24 AU in radius, well within the dust sublimation radius, and noted that this continuum may result from hot dust interior to the sublimation radius which is shielded from stellar radiation or from free-free emission from an inner gas disk \citep{m04}.  We calculated the K-band visibility of our Model B  for the three baseline orientations used by \citet{t08} and checked these calculations with the Visibility Modeling Tool provided by NExScI.  We calculated K-band visibilities of 0.99, 0.74, and 0.74 for baselines of 55 m at 34$^{\circ}$, 82 m at 91$^{\circ}$, and 121 m at 69$^{\circ}$, respectively.  These values are close to the measured visibilities of 1.0$\pm$0.1, 0.8$\pm$0.05, and 0.8$\pm$0.03; the latter two modeled visibilities fall just below the lower limit of the measured visibilities given the reported uncertainties.

Because the Keck Nuller and VLTI-MIDI do not resolve these extreme inner portions of the disk, the K-band observations may probe a different structure than what the N-band observations probe.  \citet{t08} use photometric data to estimate that the excess continuum contributes 25\% of the total flux at 2.2 microns.  By estimating the stellar flux at 2.2 microns and modeling the continuum flux source as a blackbody at 1500 K (dust at the sublimation temperature), we calculate that the continuum source's contribution at N-band would be no larger than 1.9 Jy.  This contribution is a factor of $\sim$5 less than the flux from the blackbody component of our best fit two-component optically thin disk models.  This disparity suggests that an additional blackbody component is necessary beyond what \citet{t08} model.  Additionally, our Model A and Model B blackbody components contribute only 1.4 Jy and 3.4 Jy of flux at 2.2 microns, respectively, less than the 3.6 Jy of excess continuum flux calculated by \citet{t08}.  Perhaps a three-component model, with an outer disk of small silicate grains, an inner disk of large grains, and a continuum source interior to the dust sublimation radius (shielded hot dust or free-free emission from a gas disk), will be necessary to explain all of the interferometric and spectral data sets.

\subsection{Two-layer models}

Motivated by the \citet{cg97} and \citet{d01} circumstellar disk models, we explored an alternative edge-on disk geometry consisting of a blackbody midplane layer sandwiched between two identical optically-thin surface layers of small dust grains.  The surface layers are thin; they have a very small scale height and are unresolved in the direction perpendicular to the midplane.  Our models were completely defined by 6 parameters: the disk position angle, the temperature and scale height of the middle layer, and the temperature, grain size, and line density of the surface layers.  We investigated 100 middle-layer and 100 surface-layer temperatures ranging from 300 -- 1300 K, and 6 surface layer dust sizes from 0.1 -- 1.0 $\mu\rm{m}$.  We used the surface-layer temperature, which represents the temperature of the surface layer at the outer truncation radius of the disk, to calculate the outer radius of the disk.  Because our disk is edge-on, we ignore any contribution by a hot inner wall.  We note that \citet{l04} fit the VLTI-MIDI data with a \citet{d01} flared disk model which included a hot inner wall and found this model to fit poorly. 

\begin{figure}
\begin{center}
\includegraphics[width=4.in]{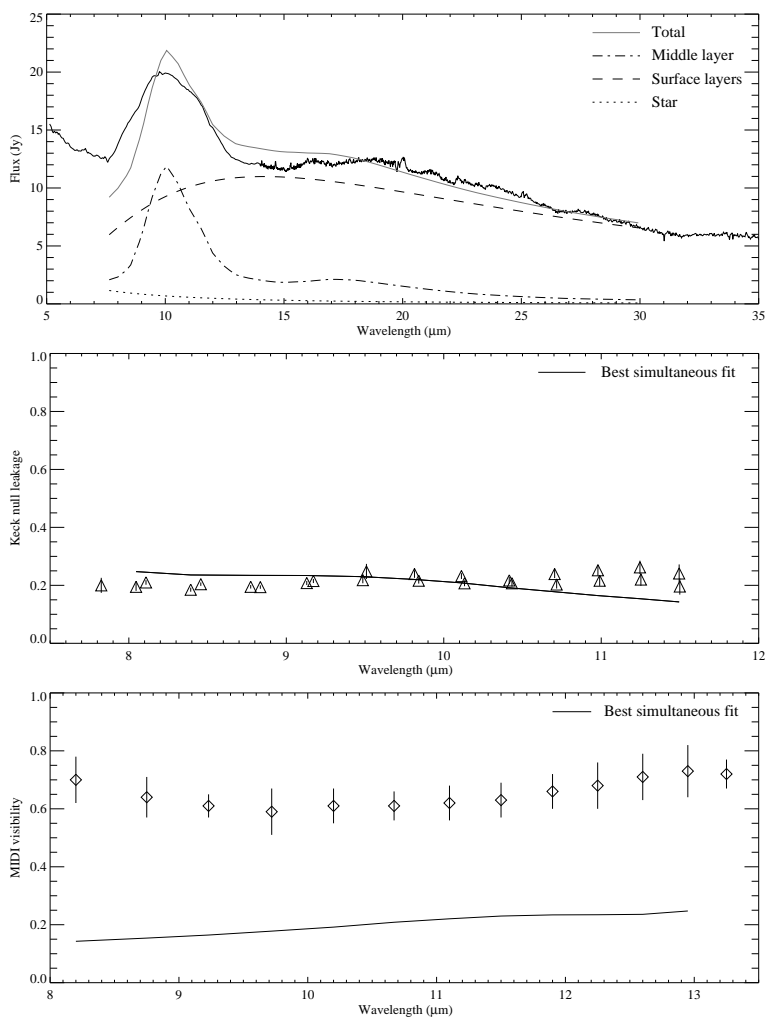}
\caption{Best simultaneous fit to the three data sets using a two-layer optically-thick disk model.  The best fit model is a disk oriented with a position angle of $38^{\circ}$ and truncated at outer radius 3.4 AU.  The disk features a middle layer of 360 K blackbody grains with a disk height of 0.163 AU from the midplane, and 2 surface layers of 0.7 $\mu\rm{m}$ grains heated to 835 K, each with line densities of $8.2 \times 10^{32}\; \rm{AU}^{-1}$.  \label{bestfit3layer_noshortlambda_figure}}
\end{center}
\end{figure}

We first derived the total emitting area of each component, and therefore the line density of the surface layers and scale height of the middle layer, by fitting the flux of the two disk components to the \emph{Spitzer} observations.  For the 500 best fits to the \emph{Spitzer} IRS spectrum, we calculated the Keck null leakage and VLTI-MIDI visibility as a function of position angle in $1^{\circ}$ increments.  Figure \ref{bestfit3layer_noshortlambda_figure} shows the single best fit to the three data sets.   The best fit two-layer model qualitatively reproduces the major features of the \emph{Spitzer} IRS spectrum, shown in the top panel of Figure \ref{bestfit3layer_noshortlambda_figure}, although it underestimates the flux near 8 $\mu\rm{m}$ and overestimates the flux from 10--18 $\mu\rm{m}$.  The lower two panels of Figure \ref{bestfit3layer_noshortlambda_figure} shows this model can reproduce the mean Keck null leakage, but cannot simultaneously reproduce the VLTI-MIDI visibility.  The interferometric data seems to rule out this disk geometry.


\section{Discussion \& Interpretation of Models \label{discussion_section}}

Our modeling efforts have yielded an optically-thin disk model that can qualitatively reproduce our 51 Oph observations with the Keck Nuller and also the VLTI-MIDI and \emph{Spitzer} observations of this system.  This model, Model B, is composed of an inner ring of large grains distributed from their sublimation radius ($\sim 0.5$ AU) out to $\sim 4$ AU, and an outer ring of small grains ($<1\; \mu\rm{m}$) distributed from an inner radius of $\sim$7 AU to an outer radius of $\sim$1200 AU.  We also experimented with optically-thick models, but found them unsatisfactory.

Our preferred optically-thin disk model could be interpreted as an inner ``birth" disk of material producing small dust grains through collisions.  Grains produced in the inner ring that are larger than the blowout size either spiral inward under Poynting-Robertson drag or are collisionally fragmented.  Grains smaller than the blowout size exit the system on hyperbolic orbits in a dynamical time; the ejected grains and marginally ejected grains correspond to the outer disk component.  The blowout size for blackbody dust grains in a disk without gas is given by
\begin{equation}
\label{blowouttimeequation}
  s_{\rm blowout} = \frac{3L_{\star}}{8\pi G M_{\star} c \rho_{\rm s}},
\end{equation}
where $L_{\star}$ and $M_{\star}$ are the luminosity and mass of the star, respectively, $G$ is the gravitational constant, $c$ is the speed of light, and $\rho_{\rm s}$ is the dust density \citep{bls79}.  Assuming $L_{\star} \approx 260\, L_{\sun}$, $M_{\star} \approx 3\, M_{\sun}$, and $\rho_{\rm s} = 2$ gm cm$^{-3}$, the blowout size for 51 Oph is approximately 50 $\mu\rm{m}$, consistent with our preferred model.  A similar pattern, a central population of large grains and an outer ring of small grains, has been seen in observations of the debris disks around $\beta$ Pictoris \citep{a01}, Vega \citep{srm05}, and AU Microscopii \citep{sc06}.

\subsection{Gas in the 51 Oph disk}

Of course the disk around 51 Oph is not gas free \citep[e.g.][]{lde97a,vdA01,b07}, so our scenario of an inner birth ring and outer disk of ejected grains requires that the gas disk is sparse enough to allow the small grains to be ejected from the system.  Here we estimate the maximum tolerable gas density for which small grains are unbound in the 51 Oph system.

The stopping time of a dust grain undergoing Epstein drag forces is given by 
\begin{equation}
  t_{\rm stop} \sim \frac{s\rho_{\rm s}}{\rho_{\rm g} c_{\rm sound}},
\end{equation}
where $\rho_{\rm g}$ is the gas density and $c_{\rm sound}$ is the sound speed \citep[see, e.g.][]{wc93}.  The dynamical timescale is given by
\begin{equation}
  t_{\rm dynamic} \sim \left(\frac{a^3}{GM_{\star}}\right)^{1/2},
\end{equation}
where $a$ is the semi-major axis of the grain.  For the dust grains to be significantly affected by gas drag, $t_{\rm stop} < t_{\rm dynamic}$.  Substituting $c_{\rm sound} = (kT_{\rm g}/m_{\rm g})^{1/2}$, where $k$ is the Boltzmann constant, $T_{\rm g}$ is the gas temperature, and $m_{\rm g}$ is the mass of an $\rm{H}_2$ molecule, we find that entraining a 1 $\mu\rm{m}$ grain in the gas would require
\begin{equation}
	\left(\frac{n_{\rm g}}{1\; \rm{cm}^{-3}}\right)	 \lesssim 10^{10}\; T_{\rm g}^{-1/2} \left(\frac{a}{1\; \rm{AU}}\right)^{-3/2}.
\end{equation}
Assuming a gas temperature of 1400 K at a circumstellar distance of 5 AU, near the vicinity of the inner disk, the gas density required to stop a 1 $\mu\rm{m}$ grain is $\sim 10^7$ cm$^{-3}$.  We conclude that if the disk around 51 Oph is comprised of an inner birth ring and an outer disk of ejected grains, then the gas density in the 51 Oph disk must be less than $\sim$100 times the gas density of $\beta$ Pic \citep{b04} at a circumstellar distance of $\sim$5 AU.

\subsection{A possible warp in the disk}

The outer disk in Model B has a scale height at least $\sim$4 times larger than the scale height of the inner disk.  So the small grains at high orbital inclinations in the outer disk seem unlikely to have originated from the thin inner disk in our simple model.  Such small grains cannot easily be perturbed to high-inclination orbits by hidden planets either, because of their short lifetimes compared to secular time scales.

The difference in scale heights between the two components of our model suggests to us that 51 Oph may have an unobserved sub-disk of larger bodies on inclined orbits in the inner regions of the disk.  The limited $(u, v)$ coverage and resolution of the Keck and VLTI-MIDI observations could easily keep such a sub-disk hidden from our observations.  This sub-disk might be analogous to the X-pattern, or ``warp" imaged by \citet{g06} in the $\beta$ Pictoris disk.  The large bodies in this sub-disk could launch the small grains onto inclined orbits, just as \citet{a01} suggested occurs in the $\beta$ Pictoris disk.

\subsection{Limitations of the model \& sources of confusion}

Here we discuss two possibilities that might complicate our interpretation of the mid-IR interferometry of 51 Oph: a) the disk composition or structure is more complex than can be represented by our models, or b) the disk has changed during the four year interval between VLTI-MIDI and Keck observations.

The chemical composition of the 51 Oph dust disk is not well-known.  Our models assumed astronomical silicates only, and ignored the possibility of a more complex composition.  A detailed model of the 51 Oph disk composition, along the lines of \citet{r08} for example, which we leave for future work, would help to further refine our models.

The disk geometry could also be more complex than our models can capture.  As previously discussed, there are a number of observations which suggest that an inner circumstellar disk is near edge-on in the 51 Oph system.  Our optically-thin models of the dust disk assume an outer disk which is coplanar, but the 51 Oph disk inclination may be more complex.  Some debris disks show warps or sub-disks at different inclinations \citep[e.g.][]{g06}.  It is possible that our assumption of coplanar disks does not allow Model A, which best fits the \emph{Spitzer} IRS spectrum, to simultaneously fit the VLTI-MIDI visibility and Keck null leakage.

The 51 Oph disk geometry may also have changed between observations.  The blowout time for small grains originating from a belt of material at 1 AU is on the order of a single dynamical time (only a few years).  So the outer disk could evolve significantly within the four years between Keck and VLTI-MIDI observations via a recent collisional event \citep[e.g.][]{vdA01}.  Additionally, 51 Oph's inner disk could feature complex resonant structures, such as clumps or rings due to the presence of planets \citep[e.g.][]{sk08}, which rotate in and out of view on a dynamical timescale.

A comparison of the \emph{Spitzer} IRS spectrum shown in Figure \ref{bestfit2comp_noshortlambda_figure} and the ISO spectrum published in \citet{vdA01} reveals that the 51 Oph disk may have changed significantly in the 8 years between spectral observations.  The silicate emission feature in the \emph{Spitzer} IRS spectrum peaks at 10 $\mu\rm{m}$, whereas the ISO spectrum peaks at 10.5 to 11 $\mu\rm{m}$.  The slope of the right side of the silicate emission feature is also noticeably steeper in the \emph{Spitzer} IRS spectrum; estimated slopes of the right side of the 10 $\mu\rm{m}$ silicate emission feature are $\sim -1.4$ Jy $\mu\rm{m}^{-1}$ and $\sim -2.9$ Jy $\mu\rm{m}^{-1}$ for ISO and \emph{Spitzer} observations, respectively.  These differences indicate that the 51 Oph disk may have had larger grains at the time of the ISO observations.  Finally, the flux longward of 20 $\mu\rm{m}$ in the ISO spectrum is $\sim 3$ Jy less than the Spitzer IRS spectrum, possibly because there were fewer cold grains at the time of the ISO observations.

\section{Summary}

We observed the 51 Oph disk at N-band using the Keck interferometer operating in nulling mode.  We combined the observed Keck null leakage with VLTI-MIDI visibility data and the \emph{Spitzer} IRS spectrum and simultaneously modeled all three data sets.  We experimented with a variety of optically-thin dust cloud models and also an edge-on optically-thick disk model.  The \emph{Spitzer} IRS spectrum ruled out the single-component optically-thin model, while the interferometric data ruled out our optically-thick model.

\begin{figure}
\begin{center}
\includegraphics[width=4.in]{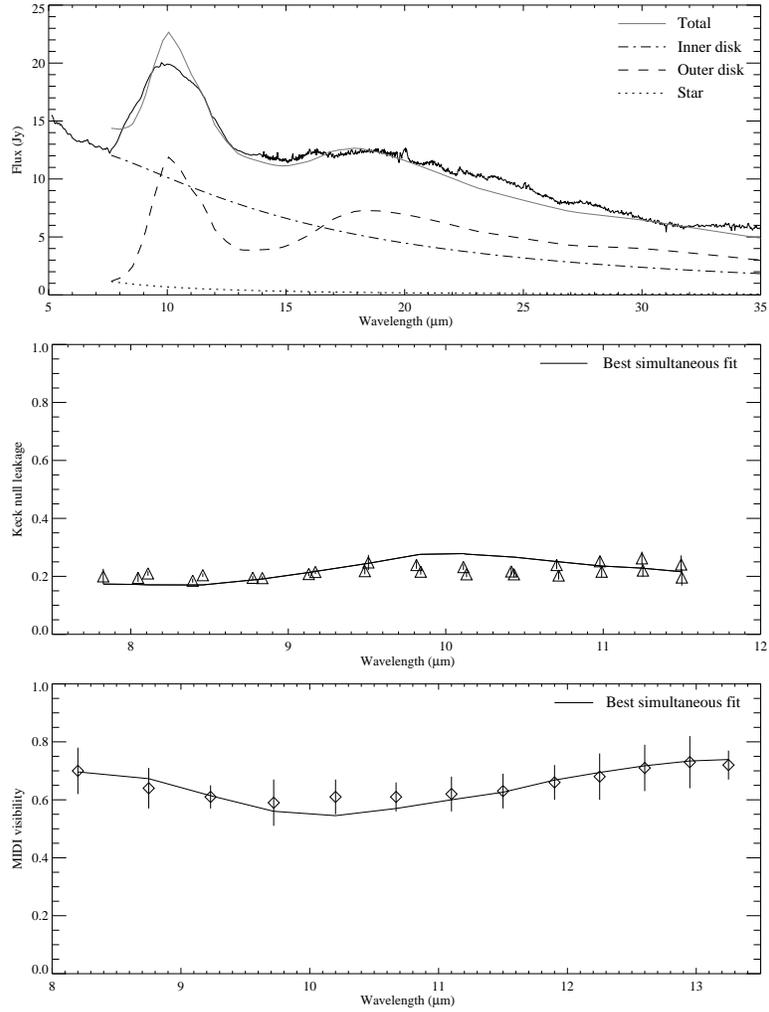}
\caption{Model B, the best fit to the three data sets using a 2-component optically thin disk model where the inner and outer radii of the inner disk are forced to 0.54 AU and 4 AU, respectively.  The best-fit parameters of Model B are listed in Table \ref{bestfitstable}.  \label{bestfit2comp_noshortlambda_figure_7}}
\end{center}
\end{figure}

Our preferred model consists of two separate populations of large and small grains.  The three data sets are best simultaneously fit by our Model B (Table \ref{bestfitstable}).  This model, shown in Figure \ref{bestfit2comp_noshortlambda_figure_7}, contains a disk of larger grains that extends from the grain sublimation radius out to $\sim 4$ AU and a disk of $0.1\; \mu\rm{m}$ grains that extends from $\sim 7$ AU to $\sim 1200$ AU.

This model may be consistent with an inner ``birth" disk of continually colliding parent bodies.  The large grains ($\gtrsim$ 50 $\mu\rm{m}$) produced by the parent bodies make up the inner disk, while the small grains ($\lesssim$ 50 $\mu\rm{m}$) are blown outward and eventually ejected from the system by radiation pressure.  The large scale height of the outer disk compared to the inner disk suggests that the small grains which compose the outer disk may originate from an unseen inclined sub-disk or from a population of inclined cometary bodies.


Although the 51 Oph disk seems puzzling at first, perhaps it is not so strange after all.  The distribution of grain sizes in our models is not unique to the 51 Oph system, but has been observed in the $\beta$ Pictoris \citep{a01}, Vega \citep{srm05}, and AU Microscopii \citep{sc06} disks.  Our models suggest that the 51 Oph disk may be another member of a class of debris disks which exhibit similar dust distributions.  Our models also indicate that there may be two sources of dust at different inclinations around 51 Oph.  These models, together with previous observations of variable absorption features \citep{gs93,r02} which may be due to transient infalling bodies suggest that the 51 Oph dust disk may well be an example of a $\beta$ Pictoris-like system.

\acknowledgments

We thank the National Aeronautics and Space Administration and Goddard Space Flight Center for support of this research through funding from the Graduate Student Researchers Program, and the Harvard-Smithsonian Center for Astrophysics and NASA's Navigator Program for their financial support via the Keck Interferometer Nuller Shared Risk Science Program.  Part of this research was carried out at the Jet Propulsion Laboratory, California Institute of Technology, under contract with the National Aeronautics and Space Administration.  The authors also thank Rachel Akeson and Rafael Millan-Gabet for their help in making these observations possible.


\newpage
\clearpage
\begin{deluxetable}{cccccc}
\tablewidth{0pt}
\footnotesize
\tablecaption{Keck Nuller Observation Log \label{observationstable}}
\tablehead{
\colhead{Object} & \colhead{Type} & \colhead{Time} & \colhead{U} & \colhead{V} & \colhead{Air Mass} \\ 
 &  & \colhead{(UT)} & \colhead{(m)} & \colhead{(m)} \\
}
\startdata
51 Oph & target & 15:08:24 & 52.00 & 48.97 & 1.39 \\
51 Oph & target & 15:09:26 & 51.90 & 48.87 & 1.39 \\
$\epsilon$ Oph & calibrator & 15:37:21 & 37.34 & 59.76 & 1.21 \\
\enddata
\vspace{-0.3in}
\end{deluxetable}

\newpage
\begin{landscape}
\clearpage
\begin{deluxetable}{lccccccccccc}
\tablewidth{0pt}
\footnotesize
\tablecaption{Best-fit optically-thin model parameters with 99.73\% joint confidence estimates \label{bestfitstable}}
\tablehead{

& \colhead{$s_1$} & \colhead{$r_{\rm in,1}$} & \colhead{$r_{\rm out,1}$} & \colhead{$h_1/r$} & \colhead{$\Sigma_1$}& \colhead{$s_2$} & \colhead{$r_{\rm in,2}$} & \colhead{$r_{\rm out,2}$} & \colhead{$h_2/r$} & \colhead{$\Sigma_2$} & \colhead{PA}\\ 

& \colhead{($\mu\rm{m}$)}  & \colhead{(AU)} & \colhead{(AU)} & & \colhead{(Zodis)\tablenotemark{c}} & \colhead{($\mu\rm{m}$)}  & \colhead{(AU)} & \colhead{(AU)} & & \colhead{(Zodis)\tablenotemark{c}} & \colhead{($^{\circ}$)}\\
}
\startdata
Model A & $0.1^{+0.05}_{-0.05}$ & $2.44^{+0.07}_{-0.0}$ & $1200^{+2300}_{-600}$ & 0.03\tablenotemark{\dag} & 1.15$^{+0.08}_{-0.08}\times$10$^5$ & 100\tablenotemark{a} & $0.716^{+0.006}_{-0.005}$ & $14.5^{+0.4}_{-0.2}$ & $0.050^{+0.004}_{-0.003}$ & 2.49$^{+0.07}_{-0.07}\times$10$^5$ & $131^{+0.15}_{-0.05}$ \vspace{+0.1in}\\

Model B & 0.1\tablenotemark{\dag} & $7.1^{+0.3}_{-0.3}$ & $1200^{+2300}_{-600}$ & 0.19\tablenotemark{\ddag} & 2.05$^{+0.06}_{-0.06}\times$10$^5$ & 100\tablenotemark{a} & 0.54\tablenotemark{b} & 4.0\tablenotemark{b} & 0.04\tablenotemark{\dag} & 4.3$^{+0.1}_{-0.1}\times$10$^5$ & $122^{+0.5}_{-0.15}$ \\
\enddata
\tablenotetext{a}{Fixed parameter in both models}
\tablenotetext{b}{Fixed value in Model B}
\tablenotetext{c}{One ``zodi" refers to a face-on optical depth of $10^{-7}$ at 1 AU}
\tablenotetext{\dag}{Upper limit}
\tablenotetext{\ddag}{{Lower limit}}
\end{deluxetable}
\end{landscape}

\newpage
\clearpage


\newpage
\clearpage

\newpage
\clearpage

\newpage
\clearpage

\newpage
\clearpage

\newpage
\clearpage

\newpage
\clearpage

\newpage
\clearpage


\begin{thebibliography}{}

\bibitem[Augereau et al.(2001)]{a01}Augereau, J. C., Nelson, R. P., Lagrange, A. M., Papaloizou, J. C. B., \& Mouillet, D. 2001, A\&A, 370, 447

\bibitem[Barry et al.(2008)]{b08}Barry, R. K., et al. 2008, \apj, 677, 1523

\bibitem[Berthoud et al.(2007)]{b07}Berthoud, M. G., Keller, L. D., Herter, T. L., Richter, M. J., \& Whelan, D. G. 2007, \apj, 660, 461


\bibitem[Brandeker et al.(2004)]{b04}Brandeker, A., Liseau, R., Olofsson, G., \& Fridlund, M. 2004, A\&A, 413, 681

\bibitem[Burns et al.(1979)]{bls79}Burns, J. A., Lamy, P. L., \& Soter, S. 1979, Icarus, 40, 1

\bibitem[Chiang \& Goldreich(1997)]{cg97}Chiang, E. I., \& Goldreich, P. 1997, ApJ, 490, 368

\bibitem[Colavita et al.(2008)]{c08}Colavita, M. M., et al. 2008, in Proc. SPIE 7013, Optical and Infrared Interferometry, ed. M. Sch\"{o}ller, W. C. Danchi, \& F. Delplancke, 70130A 



\bibitem[Doering et al.(2007)]{d07}Doering, R. L., Meixner, M., Holfeltz, S. T., Krist, J. E., Ardila, D. R., Kamp, I., Clampin, M. C., \& Lubow, S. H. 2007, \aj, 133, 2122

\bibitem[Draine \& Lee(1984)]{dl84}Draine, B. T., \& Lee, H. M. 1984, \apj, 285, 89

\bibitem[Dullemond et al.(2001)]{d01}Dullemond, C. P., Dominik, C., \& Natta, A. 2001, \apj, 560, 957

\bibitem[Dunkin et al.(1997)]{dbr97}Dunkin, S. K., Barlow, M. J., \& Ryan, S. G. 1997, MNRAS, 286, 604

\bibitem[Fajardo-Acosta et al.(1993)]{fatk93}Fajardo-Acosta, S. B., Telesco, C. M., \& Knacke, R. F. 1993, \apj, 417, 33

\bibitem[Grady \& Silvis(1993)]{gs93}Grady, C. A., \& Silvis, J. M. S. 1993, \apj, 402, 61

\bibitem[Golimowski et al.(2006)]{g06}Golimowski, D., et al. 2006, \aj, 131, 3109

\bibitem[Jayawardhana et al.(2001)]{j01}Jayawardhana, R., Fisher, R. S., Telesco, C. M., Pi\~na, R. K., Barrado y Navascu\'es, D., Hartmann, L. W., \& Fazio, G. G. 2001, \aj, 122, 2047

\bibitem[Keller et al.(2008)]{k08}Keller, L. D., et al. 2008, \apj, 684, 411

\bibitem[Kelsall et al.(1998)]{k98}Kelsall, T., et al. 1998, \apj, 508, 44

\bibitem[Lecavelier des Etangs et al.(1997)]{lde97a}Lecavelier des Etangs, A., et al. 1997, A\&A, 321, 39


\bibitem[Leinert et al.(2004)]{l04}Leinert, Ch., et al. 2004, A\&A, 423, 537


\bibitem[Monnier et al.(2009)]{m09}Monnier, J. D., et al. 2009, ApJ, submitted

\bibitem[Moran et al.(2004)]{mkh04}Moran, S. M., Kuchner, M. J., \& Holman, M. J. 2004, ApJ, 612, 1163

\bibitem[Meeus et al.(2001)]{m01}Meeus, G., Waters, L. B. F. M., Bouwman, J., van den Ancker, M. E., Waelkens, C., \& Malfait, K. 2001, A\&A, 365, 476

\bibitem[Muzerolle et al.(2004)]{m04}Muzerolle, J., D'Alessio, P., Calvet, N., \& Hartmann, L. 2004, \apj, 617, 406

\bibitem[Perryman et al.(1997)]{p97}Perryman, M. A. C., et al. 1997, A\&A, 323, L49

\bibitem[Reach et al.(2008)]{r08}Reach, W. T., Lisse, C., von Hippel, T., Mullally, F. 2008, \apj, in press

\bibitem[Roberge et al.(2002)]{r02}Roberge, A., Feldman, P. D., Lecavelier des Etangs, A., Vidal-Madjar, A., Deleuil, M., Bouret, J.-C., Ferlet, R., \& Moos, H. W. 2002, \apj, 568, 343

\bibitem[Serabyn et al.(2005)]{s05}Serabyn, E., et al. 2005, in Proc. SPIE 5905, Techniques and Instrumentation for Detection of Exoplanets II, ed. D. R. Coulter, 272

\bibitem[Stark \& Kuchner(2008)]{sk08}Stark, C. C., \& Kuchner, M. J. 2008, \apj, 686, 637

\bibitem[Strubbe \& Chiang(2006)]{sc06}Strubbe, L. E., \& Chiang, E. I. 2006, \apj, 648, 652

\bibitem[Su et al.(2005)]{srm05}Su, K. Y. L., et al. 2005, \apj, 628, 487


\bibitem[Thi et al.(2005)]{t05}Thi, W.-F., van Dalen, B., Bik, A., \& Waters, L. B. F. M. 2005, A\&A, 430, 61

\bibitem[Tatulli et al.(2008)]{t08}Tatulli, E., et al. 2008, A\&A, 489, 1151


\bibitem[van den Ancker et al.(2001)]{vdA01}van den Ancker, M. E., Meeus, G., Cami, J., Waters, L. B. F. M., \& Waelkens, C. 2001, A\&A, 369, 17

\bibitem[Waters et al.(1988)]{wcg88}Waters, L. B. F. M., Cot\'{e}, J., \& Geballe, T. R. 1988, A\&A, 203, 348

\bibitem[Weidenschilling \& Cuzzi(1993)]{wc93}Weidenschilling, S. J., \& Cuzzi, J. N. 1993, in Protostars and Planets III, ed. E. H. Levy \& J. I. Lunine, 1031
\end{thebibliography}
\end{document}